\documentclass[aps,twocolumn]{revtex4}
\usepackage{amsfonts}
\usepackage{amsmath}
\usepackage{amssymb}
\usepackage{graphicx}

\begin{document}

\title{Stability of magnetic configurations in nanorings }
\author{P. Landeros, J. Escrig and D. Altbir}
\affiliation{Departamento de F\'{\i}sica, Universidad de Santiago de Chile, USACH, Av.
Ecuador 3493, Santiago, Chile}
\author{M. Bahiana and J. d'Albuquerque e Castro}
\affiliation{Instituto de F\'{\i}sica, Universidade Federal do Rio de Janeiro, Cx.Postal
68.528, 21941-972, RJ, Brazil.}

\begin{abstract}
The relative stability of the vortex, onion and ferromagnetic phases in
nanorings is examined as a function of the ring geometry. Total energy
calculations are carried out analytically, based on simple models for each
configuration. Results are summarized by phase diagrams, which might be used
as a guide to the production of rings with specific magnetic properties.
\end{abstract}

\maketitle

\section{Introduction}

The understanding of the properties of small magnetic elements has been a
major challenge in the rapidly evolving field of nanoscale science. Besides
the basic scientific interest in the magnetic properties of these systems,
evidence is that they might be used in the production of magnetic devices,
such as high density media for magnetic recording \cite{Chou,zhu} and
magnetic logic \cite{zhu,logic}. Common structures are arrays of nanowires
\cite{vazquez3}, cylinders \cite{cylinders1,cylinders2}, cones \cite{cones1,
cones2} and rings \cite{rings1,rings2}. Recent theoretical studies on such
structures have been carried out aiming at determining the stable magnetized
state as a function of their geometric details \cite%
{Jubert,Porrati,Metlov,Guslienko}. Among the available geometries, the ring
shape is of particular interest due to its core-free magnetic configuration
leading to uniform switching fields, guaranteeing reproducibility in
read-write processes \cite{zhu}. As a consequence, magnetic nanorings have
received increasing attention over the last few years, from both
experimental and theoretical points of view.

Geometrically, ring-shaped particles are characterized by their external and
internal radii, $R$ and $a$, respectively, and height, $H$. Magnetic
measurements as well as micromagnetic simulations of such systems have
identified two in-plane magnetic states, namely the flux-closure vortex
state, $V$, and the so-called \textquotedblleft onion\textquotedblright\
state, $\mathcal{O}$. The latter is accessible from saturation and is
characterized by the presence of two opposite head to head walls \cite%
{klaui1,castano1}. In addition, for sufficiently high values of $H$, the
occurrence of ferromagnetic order, $F$, along the ring axis, is also
possible. It is therefore clear that, for practical applications, the
determination of the ranges of values of the parameters $R$, $a$, and $H$,
within which one of those configurations is of lowest energy, is of great
relevance. However, magnetic measurements do not always allow a clear
identification of the magnetic arrangement within the particles, which makes
the theoretical approach to the problem highly desirable. In the present
work we report results of total energy calculations for magnetic nanorings,
on the basis of which the relative stability of the three above mentioned
configurations could be determined. Our results are summarized by phase
diagrams in the $R-H$ plane.

\section{Total energy calculations}

Nanorings in the size range currently produced may consist of more than $%
10^{8}$ magnetic atoms. As a consequence, the determination of the
configuration of lowest energy based on the investigation of the behavior of
individual magnetic moments becomes numerically prohibitive. In order to
circumvent this problem, we resort to a simplified description of the
system, in which the discrete distribution of magnetic moments is replaced
by a continuous one, characterized by the magnetization $\overrightarrow{M}(%
\vec{r})$, such that $\overrightarrow{M}(\vec{r})\delta V $ gives the total
magnetic moment within the element of volume $\delta V$ centered at $\vec{r}$%
. We also require that $|\vec{M}(\vec{r})|=M_{0}$, the saturation
magnetization.

The internal energy, $E_{tot}$, of a single ring is given by the sum of
three terms corresponding to the magnetostatic ($E_{dip}$), the exchange ($%
E_{ex}$), and the anisotropy ($E_{K}$) contributions. Since nanorings are
usually polycrystalline, the magnetic anisotropy averages to zero due to the
random orientation of the crystallites \cite{Klaui}. In view of that, it
will be neglected in our calculations.

The dipolar contribution can be obtained from the magnetization according to
the relation

\begin{equation}
E_{dip}=\frac{\mu _{0}}{2}\int\limits_{V}\overrightarrow{M}(\vec{r})\cdot
\nabla U(\vec{r})\;\mbox{d}v\,\,,  \label{D}
\end{equation}

\noindent where an additional configuration independent term has been left
out, and the magnetostatic potential is given (in SI units) by

\begin{equation}
U(\vec{r})=-\frac{1}{4\pi }\int\limits_{V}\frac{\nabla \cdot \overrightarrow{%
M}(\vec{r}^{\prime })}{\left\vert \vec{r}-\vec{r}^{\prime }\right\vert }%
\mbox{d}v^{\prime }+\frac{1}{4\pi }\int\limits_{S}\frac{\widehat{n}^{\prime
}\cdot \overrightarrow{M}(\vec{r}^{\prime })}{\left\vert \vec{r}-\vec{r}%
^{\prime }\right\vert }\mbox{d}s^{\prime }.  \label{P}
\end{equation}

\noindent In the above expression, $V$ and $S$ represent the volume and
surface of the ring, respectively. Assuming that the magnetization varies
slowly on the scale of the lattice parameter, the exchange energy can be
written as

\begin{equation}
E_{ex}=A\int\limits_{V}\left[ \left( \nabla m_{x}\right) ^{2}+\left( \nabla
m_{y}\right) ^{2}+\left( \nabla m_{z}\right) ^{2}\right] \mbox{d}v\,\,,
\label{X}
\end{equation}

\noindent where $m_{i}=M_{i}/M_{0}$ ($i=x,y,z$) are the reduced components
of the magnetization and $A$ is the stiffness constant of the material,
which depends on the exchange interaction between the magnetic moments. Also
in this expression, an additional configuration independent term has been
left out \cite{aharoni}. At this point it is convenient to define
dimensionless variables. The natural scale for linear dimensions is given by
the exchange length, $L_{ex}=\sqrt{2A/\mu _{0}M_{0}^{2}}$, in terms of which
we can define the dimensionless radius and height $\rho =R/L_{ex}$ and $%
h=H/L_{ex}$, respectively. It is also worth defining the parameter $\gamma
=H/R$ and the aspect ratio $\beta =a/R$, such that $0<\beta <1$.

In order to proceed with our calculations, it is necessary to specify the
function $\overrightarrow{M}(\vec{r})$ corresponding to each of the three
configurations under consideration. For the ferromagnetic state, $%
\overrightarrow{M}(\vec{r})$ can be approximated by $M_{0}\hat{z}$, where $%
\hat{z}$ is the unit vector parallel to the axis of the ring, whereas for
the vortex state we can take $\overrightarrow{M}(\vec{r})=M_{0}\hat{\phi}$,
where $\hat{\phi}$ is the azimuthal unit vector in the ring plane, $xy$. The
most interesting case, however, regards the onion configuration, whose
typical arrangement of the magnetic moments is illustrated in Fig.(1). So
far, no analytical model has been presented in the literature to describe
such configuration. However in a recent work, Beleggia \textit{et al.} \cite%
{beleggia} have investigated the magnetic phase diagram for rings,
considering a ferromagnetic-in-plane configuration instead of the onion
configuration.

\begin{figure}[h]
\includegraphics[width=5cm,height=5cm]{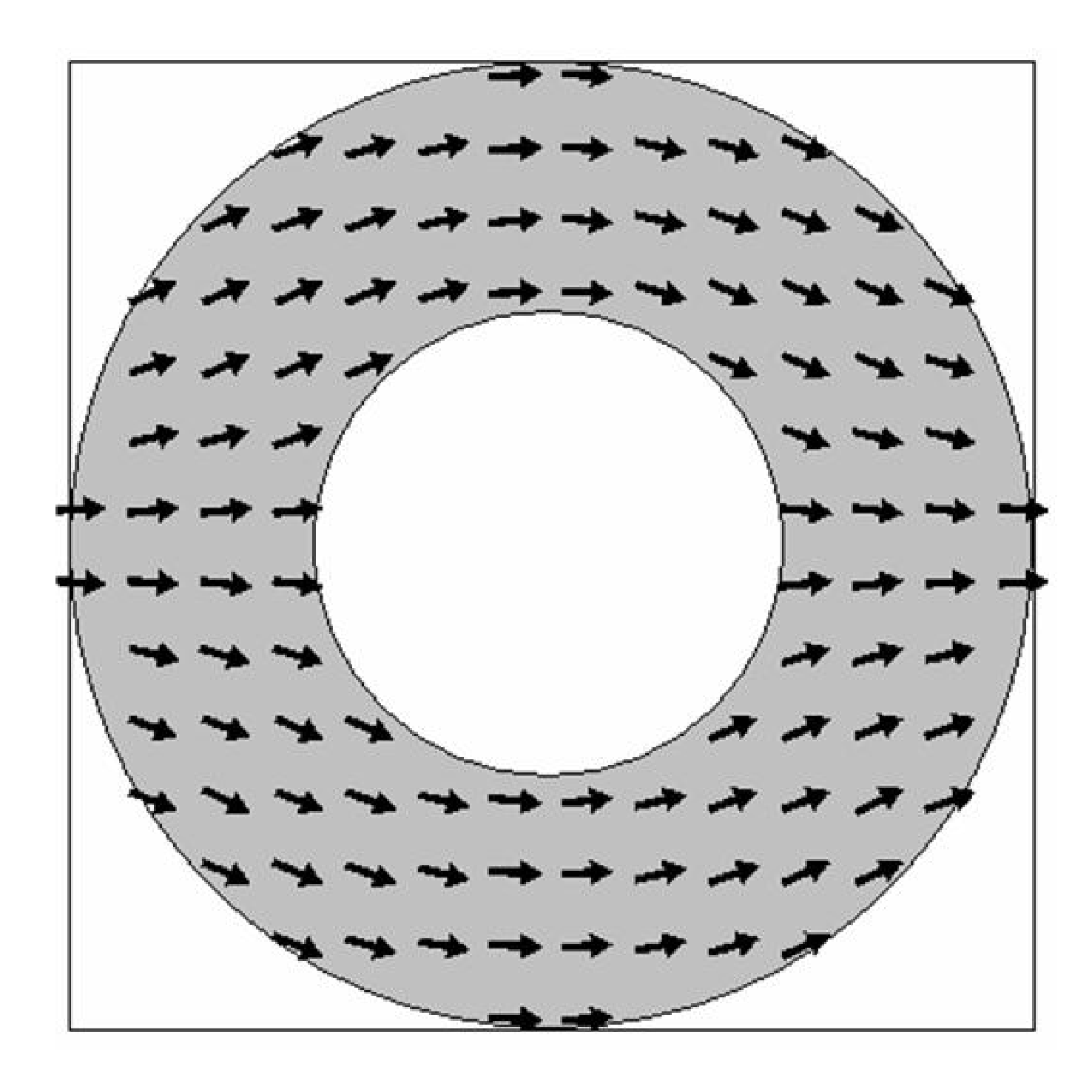} \label{f1}
\caption{Onion configuration of a Co ring with $h = 2.1$, $\protect\rho = 7$
and $\protect\beta = 0.5$ for $n=4.6$. The arrows denote the direction of
the magnetization.}
\end{figure}

The model we adopt is based on the assumption that the magnetization in the
onion configuration can be approximated by

\begin{equation}
\overrightarrow{M}(\vec{r})=M_{r}(\phi )\hat{r}+M_{\phi }(\phi )\hat{\phi}%
\,\,,  \label{MO}
\end{equation}

\noindent where the possible dependence of $M_{r}$ and of $M_{\phi }$ on $r$
has been neglected. The important point here is the implicit dependence of
the two components of the magnetization on the geometry of the ring, more
precisely, on the values of $\rho$, $h$ and $\beta $. Therefore, in order to
be realistic, any model for the onion configuration has to allow the
magnetization either to approach that of an in-plane ferromagnetic or to
deviate from it, depending on the values of these parameters. This point
will be made clearer further on in this article.

Taking into consideration the spacial symmetry of $\overrightarrow{M}$, as
represented in Fig.(1), the expressions for the reduced components $%
m_{r}=M_{r}/M_{0}$ and $m_{\phi }=M_{\phi }/M_{0}$ of the magnetization can
be written as

\begin{equation}
m_{r}(\phi )=\left\{
\begin{array}{l}
f(\phi ),\ \ \ \ 0<\phi <\pi /2 \\
-f(\pi -\phi ),\ \ \ \pi /2<\phi <\pi \\
-f(\phi -\pi ),\ \ \ \pi <\phi <3\pi /2 \\
f(2\pi -\phi ),\ \ \ 3\pi /2<\phi <2\pi \,\,,%
\end{array}%
\right.  \label{mr}
\end{equation}

\noindent and

\begin{equation}
m_{\phi }(\phi )=\left\{
\begin{array}{l}
-\sqrt{1-f^{2}(\phi )},\ \ \ \ 0<\phi <\pi /2 \\
-\sqrt{1-f^{2}(\pi -\phi )},\ \ \ \pi /2<\phi <\pi \\
\sqrt{1-f^{2}(\phi -\pi )},\ \ \ \pi <\phi <3\pi /2 \\
\sqrt{1-f^{2}(2\pi -\phi )},\ \ \ 3\pi /2<\phi <2\pi \,\,,%
\end{array}%
\right.  \label{mphi}
\end{equation}

\noindent where $f$ is a bounded function such that $-1\leq f(\phi )\leq 1$.
A suitable form for the function $f$ is
\begin{equation}
f(n,\phi )=\cos ^{n}(\phi )\,\,,  \label{model}
\end{equation}
\noindent where $n$ is a continuous variable, defined such that $n\geq 1$,
chosen so as to minimize the total energy of the configuration for each
value of $\rho $, $h$ and $\beta $. Such model allows a continuous
transition from the in-plane ferromagnetic state ($n=1$) to the onion state (%
$n>1$). The deviation from the ferromagnetic configuration becomes more
pronounced as $n$ increases from 1. Fig.(1) shows a possible onion
configuration with $n>1$.

Having specified the functional form of $\overrightarrow{M}(\vec{r})$ for
the three configurations, we are now in position to evaluate the total
energy for each of them.

\subsection{Ferromagnetic configuration ($F$)}

From Eq.(\ref{X}) we immediately find that $E_{ex}^{F}=0$. Thus, the total
energy has only the dipolar contribution, which can be obtained from the
magnetostatic potential $U(\vec{r})$ given by Eq.(\ref{P}). Using the
expansion (\ref{Green}) in expression (\ref{P}) we obtain

\begin{equation}
\tilde{E}_{tot}^{F}\equiv\frac{E_{dip}^{F}}{\mu _{0}M_{0}^{2}L_{ex}^{3}}%
=\rho^{3}\Psi (\beta ,\gamma )\,\,,  \label{EF}
\end{equation}

\noindent where the function $\Psi (\beta ,\gamma)$ is defined by

\begin{equation}
\Psi (\beta ,\gamma )\equiv \pi \int\limits_{0}^{\infty }\frac{1-e^{-\gamma
y}}{y^{2}}\left[ J_{1}(y)-\beta J_{1}(\beta y)\right] ^{2}\;\mbox{d}y\,\,.
\label{Fi}
\end{equation}
\noindent Here $J_{1}(z)$ are Bessel functions of the first kind.

Equation (\ref{EF}) has been previously obtained by Beleggia \textit{et al.}
(See Eq.(15), Ref. \cite{beleggia}) considering a more general approach
based on Fourier Transforms of the magnetization.

\subsection{Vortex configuration ($V$)}

It is clear from Eq.(\ref{P}) that for this configuration $E_{dip}^{V}=0$.
The exchange contribution can also be evaluated from Eq.(\ref{X}) so that
the final result for the reduced total energy reads
\begin{equation}
\tilde{E}_{tot}^{V}\equiv\frac{E_{ex}^{V}}{\mu _{0}M_{0}^{2}L_{ex}^{3}} =
-\pi h\ln \beta \,\,.  \label{Ev}
\end{equation}

\subsection{Onion configuration ($\mathcal{O}$)}

In the previous two configurations, the magnetization function satisfies the
condition $\nabla \cdot \overrightarrow{M}=0$, which means absence of
volumetric magnetic charges. In the onion configuration, however, due to the
presence of two regions with head-to-head domain walls, in which $\nabla
\cdot \overrightarrow{M}\neq 0$, the dipolar energy turns out to be given by
the sum of two contributions, $E_{dS}^{\mathcal{O}}$ and $E_{dV}^{\mathcal{O}%
}$, coming from the surface and volume terms in the expression for $U(\vec{r}%
)$, respectively. Details of the calculations of these two quantities are
given in the Appendix. Results for the reduced energies read

\begin{equation}
\tilde{E}_{dS}^{\mathcal{O}}\equiv \frac{E_{dS}^{\mathcal{O}}}{\mu
_{0}M_{0}^{2}L_{ex}^{3}}=\rho ^{3}\sum_{p=1}^{\infty }b_{p}\left(
b_{p}Q_{1}+d_{p}Q_{2}\right) \,,  \label{eds}
\end{equation}

\begin{equation}
\tilde{E}_{dV}^{\mathcal{O}}\equiv \frac{E_{dV}^{\mathcal{O}}}{\mu
_{0}M_{0}^{2}L_{ex}^{3}}=\rho ^{3}\sum_{p=1}^{\infty }(d_{p}-b_{p})\left(
b_{p}Q_{3}+d_{p}Q_{4}\right) \,,  \label{edv}
\end{equation}%
\noindent where

\begin{eqnarray}
b_{p}(n) &=&\int\limits_{0}^{\pi /2}f(n,\phi )\cos (p\phi )\;\mbox{d}\phi
\,\,,  \label{bm} \\
d_{p}(n) &=&p\int\limits_{0}^{\pi /2}\sqrt{1-f^{2}(n,\phi )}\sin (p\phi )\;%
\mbox{d}\phi \,\,,
\end{eqnarray}

\noindent and
\begin{multline*}
Q_{1}(\beta ,\gamma )=\frac{4}{\pi }\int\limits_{\beta }^{1}x\;\mbox{d}%
x\int\limits_{0}^{\infty }\mbox{d}y\left[ J_{p}(y)-\beta J_{p}(\beta y)%
\right] \\
\left( \frac{e^{-\gamma y}+\gamma y-1}{y}\right) \left[
J_{p-1}(xy)-J_{p+1}(xy)\right] \,\,,
\end{multline*}%
\begin{multline*}
Q_{2}(\beta ,\gamma )=\frac{8}{\pi }\int\limits_{\beta }^{1}\mbox{d}%
x\int\limits_{0}^{\infty }\mbox{d}y\left[ J_{p}(y)-\beta J_{p}(\beta y)%
\right] \\
\left( \frac{e^{-\gamma y}+\gamma y-1}{y^{2}}\right) J_{p}(xy)\,\,,
\end{multline*}%
\begin{multline*}
Q_{3}(\beta ,\gamma )=\frac{4}{\pi }\int\limits_{\beta }^{1}\mbox{d}%
w\int\limits_{\beta }^{1}x\;\mbox{d}x\int\limits_{0}^{\infty }\mbox{d}y\left[
J_{p-1}(xy)-J_{p+1}(xy)\right] \\
\left( \frac{e^{-\gamma y}+\gamma y-1}{y}\right) J_{p}(wy)\,,
\end{multline*}%
\begin{multline*}
Q_{4}(\beta ,\gamma )=\frac{8}{\pi }\int\limits_{\beta }^{1}\mbox{d}%
w\int\limits_{\beta }^{1}\mbox{d}x\int\limits_{0}^{\infty }\mbox{d}%
y\;J_{p}(wy) \\
\left( \frac{e^{-\gamma y}+\gamma y-1}{y^{2}}\right) J_{p}(xy)\,\,.
\end{multline*}%
\noindent We remark that the sum in Eqs. (\ref{eds}) and (\ref{edv}) runs
over odd values of $p$ and converges quite rapidly so that, in practice,
just a few values of $p$ (say, 5 or 6) have to be considered.

The exchange energy can be also obtained analytically, as indicated in the
Appendix. The reduced energy $\tilde{E}_{ex}^{\mathcal{O}}$ is then given by
\begin{equation}
\tilde{E}_{ex}^{\mathcal{O}}\equiv \frac{E_{ex}^{\mathcal{O}}}{\mu
_{0}M_{0}^{2}L_{ex}^{3}} =-\pi h\left[ I(n)-1\right] \ln \beta \,,
\label{exo}
\end{equation}
\noindent where
\begin{equation*}
I(n)=\frac{2}{\pi }\int\limits_{0}^{\pi /2}\frac{1}{1-f^{2}(n,\phi )}\left[
\frac{\partial f(n,\phi )}{\partial \phi }\right] ^{2}\;\mbox{d}\phi \,.
\end{equation*}

As already pointed out, the value of $n$ in the above expressions is chosen
so as to minimize the total energy, for fixed values of the geometrical
parameters $\beta $, $\gamma $ and $\rho$. In other words, it is obtained by
solving the equation $\partial \tilde{E}_{tot}^{\mathcal{O}}/\partial n=0$,
where $\tilde{E}_{tot}^{\mathcal{O}}=\tilde{E}_{dS}^{\mathcal{O}}+\tilde{E}%
_{dV}^{\mathcal{O}}+\tilde{E}_{ex}^{\mathcal{O}}$. It is worth pointing out
that the dependence of $\tilde{E}_{tot}^{\mathcal{O}}$ on $n$ enters just
via the functions $b_{p}(n)$, $d_{p}(n)$ and $I(n)$. The first one can be
expressed in terms of Gamma functions, whereas the remaining two can be
easily evaluated numerically. On the other hand, the functions $Q_{j}$ can
be calculated just once for given values of $\beta $ and $\gamma .$

\section{Results and Discussion}

The above expressions for the total energy enable us to
investigate the relative stability of the three configurations of
interest. For each aspect ratio $\beta $ we can determine the
ranges of values of the dimensionless radius $\rho $ and height
$h$ within which one of the three configurations is of lowest
energy. The boundary line between any two configurations can be
obtained by equating the expressions for the corresponding total
energies. Figure (2) illustrates phase diagrams for different
values of $\beta $. The stability regions corresponding to each
configuration are indicated in the diagram.
\bigskip

\begin{figure}[h]
\includegraphics[width=7cm,height=6cm]{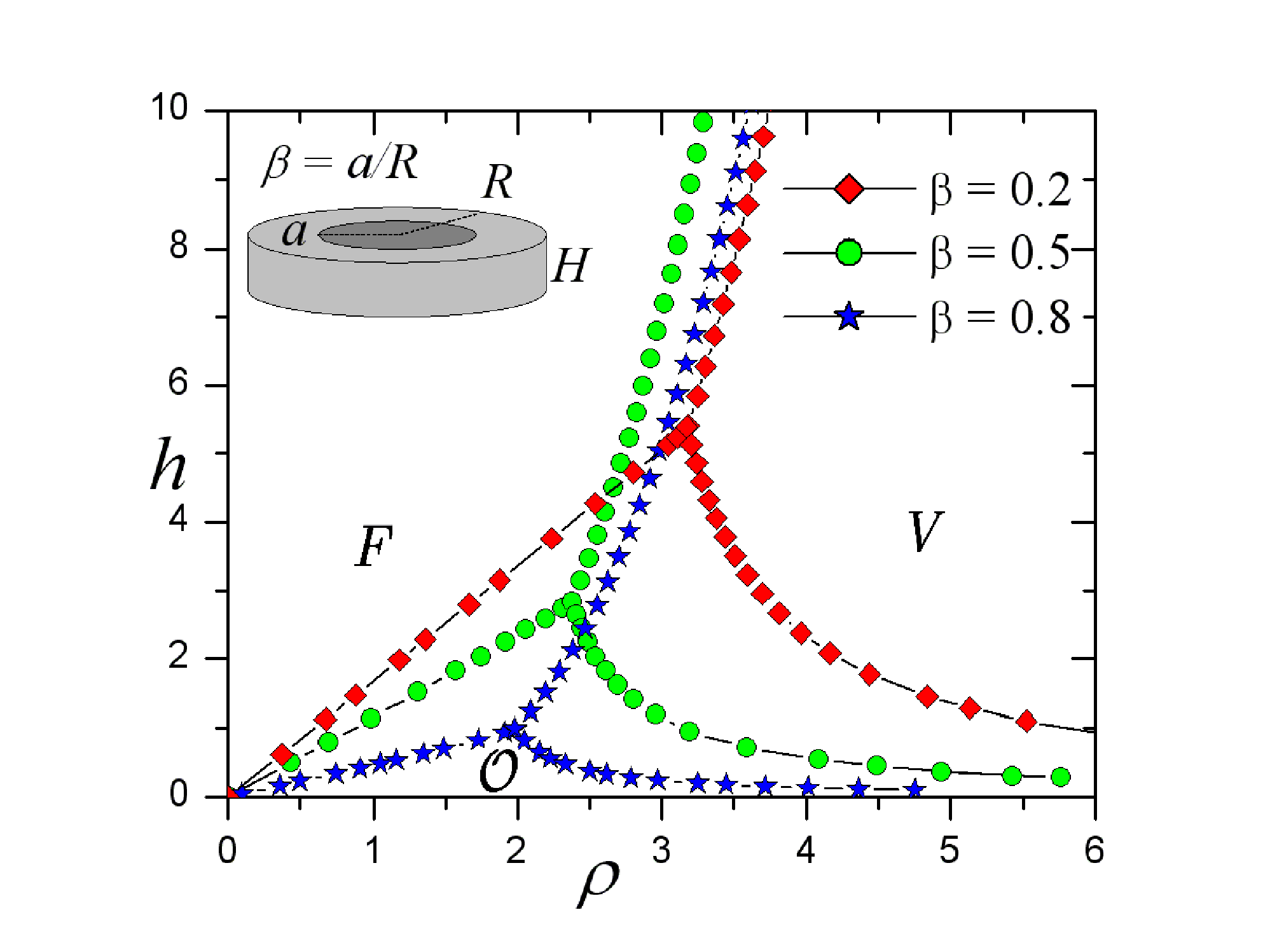} \label{f2}
\caption{Magnetic phase diagrams for rings with $\protect\beta =a/R=0.2$
(diamonds), $0.5$ (full circles), and $0.8$ (stars). $F$, $V$ and $\mathcal{O%
}$ indicate regions where ferromagnetic out-of-plane, vortex and onion
configurations are relatively more stable. $h$ and $\protect\rho$ are the
dimensionless geometric parameters, defined respect to the material exchange
length as $h=H/L_{ex}$ and $\protect\rho =R/L_{ex}$. The inset illustrates
the geometry of a ring.}
\end{figure}

It is interesting to look at the dependence of the exponent $n$, which
determines the magnetization profile in the lower energy onion
configuration, on the geometry of the ring. Fig.(3) presents a set of lines
in the $\rho h$-plane corresponding to constant values of $n$, for $\beta
=0.5$. The points on such lines laying outside the $\mathcal{O}$ region
(open symbols) correspond to either metastable or unstable onion
configurations. We shall come back to this point further on in this article.
We clearly see that for small rings, the onion configuration turns out to be
rather close to the in-plane ferromagnetic one ($n=1$). Another interesting
point is the weak dependence of the in-plane magnetization of the onion
configuration as a function of $n$. The reduced magnetization $\overline{m}%
_{x}$ can be obtained by integrating $M_{x}$ within the volume of the ring,
which gives us $\overline{m}_{x}=\left( 2/\pi \right) \left[ {b}_{{1}}{(n)+d}%
_{{1}}{(n)}\right] $. The inset in Fig.(3) presents $\overline{m}_{x}$\ as a
function of $n$, which well illustrates this point.

\begin{figure}[h]
\includegraphics[width=7cm,height=6cm]{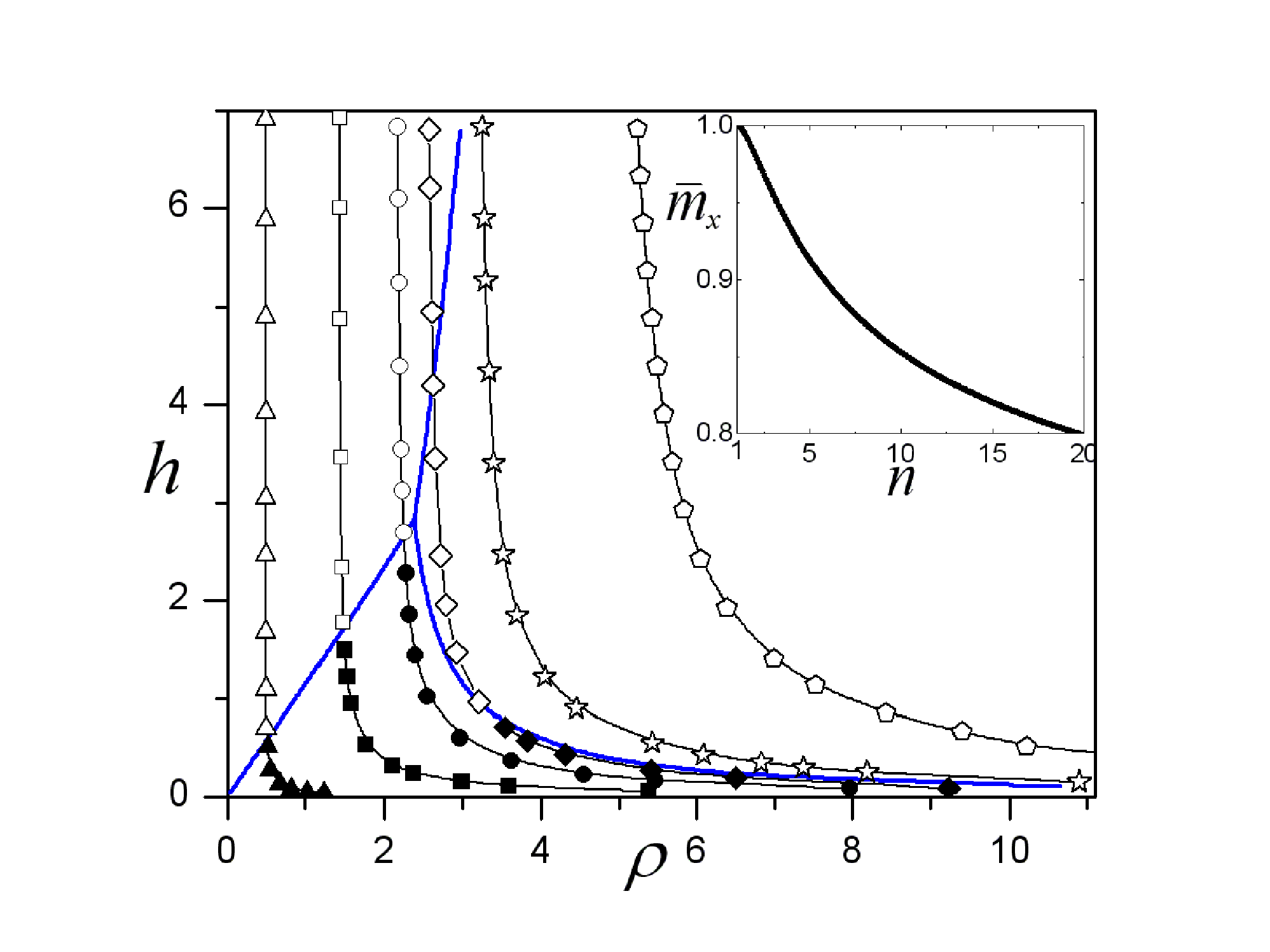} \label{f3}
\caption{Constant $n$ curves for rings with $\protect\beta =0.5$. Solid
symbols correspond to stable onion configurations, and open symbols to
metastable or unstable ones. Thick solid lines denote the boundaries between
stability regions and thin lines are guides to the eye. Results are
presented for $n=1.01$ (triangles), $n=1.1$ (squares), $1.3$ (circles), 1.5
(diamonds), 2 (stars), and 4 (pentagons). Inset: magnetization component
along the onion axis ($m_{x}$) as a function of constant-$n$ curves in the $%
\protect\rho h$-plane for rings with $\protect\beta =0.5$. Results are
presented for rings both inside (full symbols) and outside (open symbols)
the onion stability region. The latter correspond to either unstable or
metastable onion configurations.}
\end{figure}

The transition between the $F$ and $\mathcal{O}$ configurations is
determined by the balance between the energies of the out-of-plane and of
the in-plane magnetic configurations. A qualitative understanding of such
transition can be achieved by examining the strength of the demagnetization
fields along the $z$ and $x$ axes. For given $\beta $ and $\rho $, larger
values of $h$ result in reduced demagnetization fields along the $z$
direction, favoring the $F$ configuration. This is the case, for example, of
rings with $\rho =1.0$ and $\beta =0.2$, when $h$ is increased from 1.0 to
3.0. On the other hand, for fixed $\rho $ and $h$, a decrease in $\beta $ is
equivalent to a reduction of the inner radius, which results in a larger in
plane demagnetizing field. Thus, a ring with $\rho =1.0$ and $h=1.0$
exhibits an in-plane magnetic order for $\beta =0.2$, and an out-of-plane
order for $\beta =0.8$.

Concerning the two in-plane configurations, namely $\mathcal{O}$ and $V$,
the transition between them depends on the strength of the demagnetization
field along the $x$ direction in the domain wall regions of the $\mathcal{O}$
configuration. The larger the value of the ring width $W=R-a$, the smaller
this field is, favoring the onion state. Thus, a ring with $\rho =3.0$ and $%
h=1.0$ exhibits the $V$ phase for $\beta =0.8$, and the $\mathcal{O}$
configuration for $\beta =0.2$.

The mechanism responsible for the transition between the $V$ and $F$
configurations is somewhat subtle. We notice in Fig.(2) that the line
separating these two regions exhibits a maximum shift to the left for $\beta
$ about 0.5. As a consequence, rings with say $\rho =3.0$, $h=6.0$ and inner
radius corresponding to $\beta =$ 0.2, 0.5, and 0.8 are found to exhibit $F$%
, $V$, and $F$ configurations, respectively. Such interesting behavior can
be understood by considering the relative strength of the exchange and
dipolar energies in the two states. It is clear from Eq.(\ref{Ev}) that for
fixed values of \ $\rho $ and $h$ the exchange energy in the vortex state
diverges at $\beta =0$, whereas that for the $F$ configuration approaches a
finite value [cf. Eqs(\ref{EF}) and (\ref{Fi})]. Thus, for sufficiently
small inner radii ($\beta \ll 1$), the $F$ configuration has lower energy
than the $V$ one. The reason for that is the large contribution to $%
E_{ex}^{V}$ coming from the central region of the ring. Then, the increase
in the inner radius leads to a rapid decrease of $E_{ex}^{V}$, reducing $%
E_{tot}^{V}$ with respect to $E_{tot}^{F}$ and favoring the $V$
configuration for intermediate values of $\beta $. However, for values of $%
\beta $ close to 1 (narrow rings) and sufficiently large values of $h$, the
dipolar energy in the $F$ state, which can be associated to the magnetic
charges at the top and bottom surfaces of the ring, becomes smaller than $%
E_{ex}^{V}$, favoring the ferromagnetic state.

\bigskip

\begin{figure}[h]
\includegraphics[width=6cm,height=5cm]{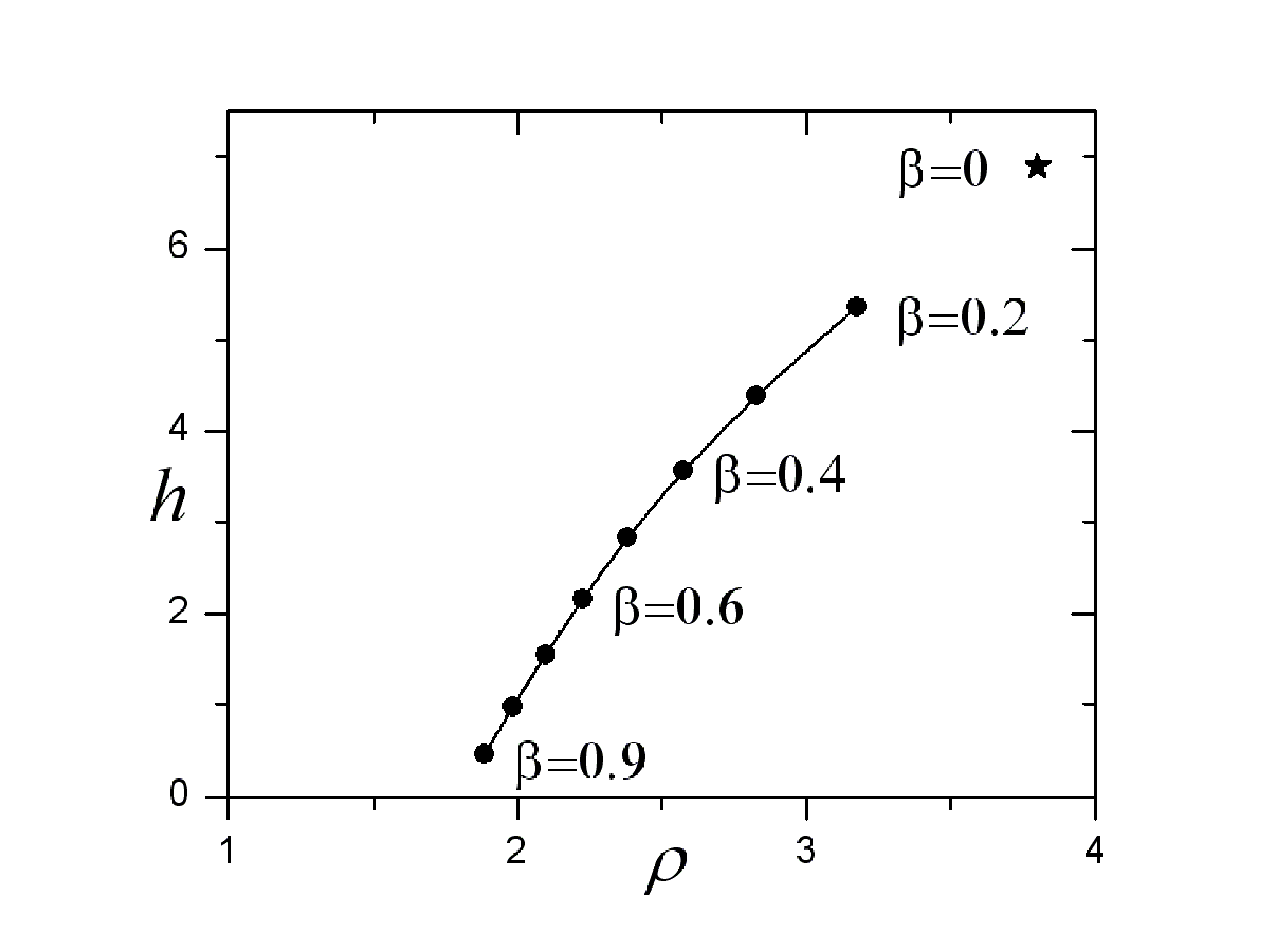} \label{f4}
\caption{Triple point position ($\protect\rho _{t},h_{t}$) as a function of $%
\protect\beta =a/R$ (full circles). The result for a cylinder is represented
by a star.}
\end{figure}

All diagrams in Fig.(2) show a triple point ($\rho_{t},h_{t}$),
corresponding to the situation in which the three configurations have equal
energy. Fig.(4) shows the trajectory of the triple point ($r_{t}(\beta
),h_{t}(\beta )$) as a function of $\beta $. We notice that the curve
correctly extrapolates to the result for cylinders, which has been
numerically obtained by d'Albuquerque e Castro \textit{et al.} \cite%
{scaling1} using a scaling technique, and analytically, by Landeros \textit{%
et al.} \cite{landerosscaling}.

An interesting result is the occurrence in some cases of
metastable onion configurations. Casta\~{n}o \textit{et al.}
\cite{castano1} have carried out
a detailed experimental study of the magnetic behavior of Co nanorings with $%
H=10$ nm and external and inner radii ranging from $90$ nm to $260$ nm and
from $20$ nm to $140$ nm, respectively. For rings with $R=180$ nm and $a=20$
nm, and $R=180$ nm and $a=70$ nm, they have found that the corresponding
hysteresis curves clearly indicate the occurrence of metastable onion
configurations. Such states can be reached by reducing the applied magnetic
field from the saturation value to zero. However, as the magnetic field is
increased in the opposite direction and the magnetic energy becomes
sufficiently high to overcome the energy barrier between the $\mathcal{O}$
and $V$ configurations, the systems undergoes a transition to the latter
state. Such behavior is entirely consistent with our results. Indeed, taking
into consideration that for Co $L_{ex}=2.85$ nm, the rings examined by Casta%
\~{n}o and collaborators have reduced dimensions $\rho = 63.2$, and $h =3.5$
and correspond to $\beta = 0.11$ and 0.39. Thus, from the results in Fig.
(2), we expect the two rings to fall well within the vortex region. We
remark that from Fig. (4), the upper bound for the position of the triple
point for $\beta \rightarrow 0$, is $\rho_{max} = 3.8$ which is much smaller
than the value corresponding to the two samples. As a consequence, the
observed onion configurations in such systems represent necessarily
metastable states. The occurrence of metastable onion states within the
vortex region has also been found in micromagnetic calculations. Using the
OOMMF package \cite{OOMMF} we have carried out ground state calculations for
a Co ring with $h=2.1$, $\rho=7$, and $\beta =0.5$. We have found that the
system might be trapped in a metastable onion configuration when the
in-plane ferromagnetic state is taken as the starting configuration. These
results are presented in Fig.(1).

In conclusion, we have theoretically investigated the dependence of the
internal magnetic configuration of nanorings on their geometry. Three
typical configurations have been considered, namely out-of-plane
ferromagnetic, vortex, and the so-called onion configuration. For the
latter, we have proposed a simple analytical model, which allows a continuum
transition between the onion and the in-plane ferromagnetic states. Our
results are summarized in phase diagrams giving the relative stability of
the three configurations. The possibility of the systems assuming unstable
or metastable configurations has also been considered. Our results might be
used as a guide to experimentalists interested in producing samples with
specific magnetic properties.

\section*{Acknowledgments}

This work has been partially supported by Fondo Nacional de Investigaciones
Cient\'{\i}ficas y Tecnol\'{o}gicas (FONDECYT, Chile) under Grants Nos.
1050013 and 7050273, and Millennium Science Nucleus "Condensed Matter
Physics" P02-054F of Chile, and CNPq, FAPERJ, PROSUL Program, and Instituto
de Nanoci\^{e}ncias/MCT of Brazil. CONICYT Ph.D. Program Fellowships,
MECESUP USA0108 project and Graduate Direction of Universidad de Santiago de
Chile are also acknowledged. P. L. and J. E. are grateful to the Physics
Institute of Universidade Federal do Rio de Janeiro for hospitality.

\section{Appendix}

The calculation of the dipolar energy of the onion configuration begins by
replacing the functional form Eq.~(\ref{MO}) in Eq.~(\ref{D}), leading to
\begin{equation}
E_{dip}^{\mathcal{O}}=\frac{\mu _{0}}{2}\int\limits_{V}\left[ M_{r}\left(
\phi \right) \frac{\partial U}{\partial r}+\frac{M_{\phi }\left( \phi
\right) }{r}\frac{\partial U}{\partial \phi }\right] \mbox{d}v\,.
\label{energia}
\end{equation}%
For the calculation of the magnetostatic potential we use expression (\ref%
{MO}) and the following expansion \cite{Jackson}
\begin{equation}
\frac{1}{\left\vert \vec{r}-\vec{r}^{\prime }\right\vert }=\sum_{p=-\infty
}^{\infty }e^{ip(\phi -\phi ^{\prime })}\int\limits_{0}^{\infty
}J_{p}(kr)J_{p}(kr^{\prime })e^{k(z_{<}-z_{>})}\mbox{d}k,  \label{Green}
\end{equation}%
where $J_{p}$ are the first kind Bessel functions. This way the surface
contribution to the potential\ (second term in Eq.(\ref{P})), reads
\begin{multline*}
U_{S}=\frac{M_{0}}{4\pi }\sum_{p=-\infty }^{\infty }e^{ip\phi
}\int\limits_{0}^{2\pi }m_{r}\left( \phi ^{\prime }\right) e^{-ip\phi
^{\prime }}\mbox{d}\phi ^{\prime } \\
\int\limits_{0}^{\infty }\mbox{d}k\;J_{p}(kr)\left[ RJ_{p}\left( kR\right)
-aJ_{p}(ka)\right] \int\limits_{0}^{H}e^{k\left( z_{<}-z_{>}\right) }\;%
\mbox{d}z^{\prime }\,.
\end{multline*}%
Using $m_{r}\left( \phi \right) $\ defined by Eq. (\ref{mr}) it\ is
straightforward to obtain
\begin{equation*}
\int\limits_{0}^{2\pi }m_{r}\left( \phi \right) e^{-ip\phi }\;\mbox{d}\phi
=2\left( 1-e^{-ip\pi }\right) b_{p}\,,
\end{equation*}%
where $b_{p}$ is defined by Eq. (\ref{bm}). This relation lead us to
consider only odd values of the sum index $p$ in what follows. Integrating
over $z^{\prime }$, and after some manipulations, we obtain that
\begin{eqnarray*}
U_{S} &=&\frac{2M_{0}}{\pi }\sum_{p=1}^{\infty }b_{p}\cos (p\phi
)\!\!\int\limits_{0}^{\infty }\mbox{d}k\;J_{p}(kr) \\
&&\times \left[ RJ_{p}\left( kR\right) -aJ_{p}(ka)\right] \left[ \frac{%
2-e^{-kz}-e^{-k(H-z)}}{k}\right] \,.
\end{eqnarray*}%
Introducing this potential in Eq. (\ref{energia}) we obtain%
\begin{eqnarray}
&&E_{dS}^{\mathcal{O}}=\frac{\mu _{0}M_{0}^{2}}{\pi }\sum_{p=1}^{\infty
}b_{p}\int\limits_{0}^{\infty }\frac{\mbox{d}k}{k}\int\limits_{0}^{H}\mbox{d}%
z\left[ 2-e^{-kz}-e^{-k(H-z)}\right]  \notag \\
&\times &\!\!\!\!\!\!\left. \int\limits_{a}^{R}\mbox{d}r\left[
RJ_{p}(kR)-aJ_{p}(ka)\right] \right\{ \frac{kr}{2}\left[
J_{p-1}(kr)-J_{p+1}(kr)\right]  \notag \\
&\times &\!\!\left. \!\!\!\!\int\limits_{0}^{2\pi }\mbox{d}\phi \;\left[
m_{r}(\phi )\cos (p\phi )-J_{p}(kr)p\,m_{\phi }(\phi )\sin (p\phi )\right]
\right\} \,.  \label{Eds2}
\end{eqnarray}
Using
\begin{equation*}
\int\limits_{0}^{2\pi }m_{r}\left( \phi \right) \cos (p\phi )\;\mbox{d}\phi
=4b_{p}
\end{equation*}%
and
\begin{equation*}
p\int\limits_{0}^{2\pi }m_{\phi }\left( \phi \right) \sin (p\phi )\;\mbox{d}%
\phi =-4d_{p}\,,
\end{equation*}%
in Eq. (\ref{Eds2})\ we obtain
\begin{eqnarray*}
E_{dS}^{\mathcal{O}} &=&\mu _{0}M_{0}^{2}R^{3}\sum_{p=1}^{\infty }\frac{8}{%
\pi }\int\limits_{0}^{\infty }\mbox{d}y \\
&&\left[ J_{p}\left( y\right) -\beta J_{p}(\beta y)\right] \left( \frac{%
e^{-\gamma y}+\gamma y-1}{y^{2}}\right) \\
&&\int\limits_{\beta }^{1}\mbox{d}x\left\{ b_{p}d_{p}J_{p}(xy)+b_{p}^{2}%
\frac{xy}{2}\left[ J_{p-1}(xy)-J_{p+1}(xy)\right] \right\} \,.
\end{eqnarray*}%
Thus, the superficial contribution to the reduced dipolar energy (Eq. (\ref%
{eds})) can be written as
\begin{equation*}
\tilde{E}_{dS}^{\mathcal{O}}\equiv \frac{E_{dS}^{\mathcal{O}}}{\mu
_{0}M_{0}^{2}L_{ex}^{3}}=\rho ^{3}\sum_{p=1}^{\infty }b_{p}\left(
b_{p}Q_{1}+d_{p}Q_{2}\right) \,.
\end{equation*}%
The calculation of the volumetric contribution (Eq. (\ref{edv})) follow the
same procedure.

The exchange energy of the onion configuration comes from substituting the
Cartesian magnetization components
\begin{eqnarray*}
m_{x}\left( \phi \right) &=&m_{r}\left( \phi \right) \cos (\phi )-m_{\phi
}\left( \phi \right) \sin (\phi ) \\
m_{y}\left( \phi \right) &=&m_{r}\left( \phi \right) \sin (\phi )+m_{\phi
}\left( \phi \right) \cos (\phi )\,,
\end{eqnarray*}%
in the semiclassical expression Eq. (\ref{X}), from which we obtain that
\begin{equation*}
\tilde{E}_{ex}^{\mathcal{O}}=\frac{h}{2}\log \frac{1}{\beta }%
\int\limits_{0}^{2\pi }\left[ \left( \frac{\partial m_{r}}{\partial \phi }%
-m_{\phi }\right) ^{2}+\left( \frac{\partial m_{\phi }}{\partial \phi }%
+m_{r}\right) ^{2}\right] \mbox{d}\phi \,.
\end{equation*}%
By using Eqs. (\ref{mr}) and (\ref{mphi}) we found the reduced exchange
energy of the onion configuration (Eq. \ref{exo}).


\begin{thebibliography}{99}
\bibitem{OOMMF} The public domain package is available at
math.nist.gov/oommf.
\end{thebibliography}
\end{document}